\documentclass[aps,prl,twocolumn,showpacs,amsmath,amssymb,superscriptaddress]{revtex4}

\usepackage{graphicx}
\usepackage{dcolumn}
\usepackage{bm}

\begin{document}


\title{Short-range spectroscopic ruler based on a single-molecule 
optical switch}




\author{Mark Bates}
\altaffiliation{These authors contributed equally to this work.}
\affiliation{Division of Engineering and Applied Sciences, Harvard University, Cambridge, Massachusetts 02138}
\affiliation{Department of Chemistry, Harvard University, Cambridge, Massachusetts 02138}

\author{Timothy R. Blosser}
\altaffiliation{These authors contributed equally to this work.}
\affiliation{Graduate Program in Biophysics, Harvard University, Cambridge, Massachusetts 02138}
\affiliation{Department of Chemistry, Harvard University, Cambridge, Massachusetts 02138}

\author{Xiaowei Zhuang}
\email[To whom correspondence should be addressed.\\E-mail: ] {zhuang@chemistry.harvard.edu}
\affiliation{Department of Chemistry, Harvard University, Cambridge, Massachusetts 02138}
\affiliation{Department of Physics, Harvard University, Cambridge, Massachusetts 02138}


\date{\today}

\begin{abstract}
We demonstrate a novel all-optical switch consisting of two molecules: a primary fluorophore that can be switched between a fluorescent and a dark state by light of different wavelengths, and a secondary chromophore that facilitates switching. The interaction between the two molecules exhibits a distance dependence much steeper than that of F\"{o}rster resonance energy transfer.  This enables the switch to act as a ruler with the capability to probe distances difficult to access by other spectroscopic methods, thus presenting a new tool for the study of biomolecules at the single-molecule level.
\end{abstract}


\maketitle

Single-molecule technologies have made a significant impact in many areas of science and engineering, ranging from electronics and photonics to molecular and cell biology. The detection of single biomolecules has transformed the way we study biological systems \cite{neher76,block90,lu98,bustamante03,yildiz03,zhuang00,weiss00}. It is now evident that proteins, RNA and even DNA molecules are much more dynamic than previously thought, exhibiting structural dynamics essential for their function. Experiments that probe the behaviour of individual molecules in real time have proven to be ideal for characterizing these structural dynamics and their functional implications.

F\"{o}rster resonance energy transfer (FRET) is the most widely adopted technique for studying the conformational dynamics of individual biomolecules \cite{zhuang00,weiss00,selvin95,ha96}. The FRET efficiency $\epsilon$ between a donor and an acceptor fluorophore depends on the intermolecular distance $R$, with $\epsilon = 1 / (1 + (R / R_0)^6)$. The F\"{o}rster radius $R_0$ is typically 4 - 6 nm for donor-acceptor pairs that are sufficiently bright to be detected at the single-molecule level. Therefore, single-molecule FRET is a useful probe of conformational changes on the order of several nanometers \cite{zhuang00,weiss00,selvin95,ha96,zhuang02,schuler02,mckinney03}.  Few methods exist for the study of structural dynamics at shorter length scales. A promising technique for monitoring angstrom-scale distances, based on electron transfer, has recently been demonstrated using a flavin enzyme \cite{yang03}.  However, the length scale of 1 - 3 nm remains difficult to access by non-invasive single-molecule techniques.

\begin{figure*}
\includegraphics{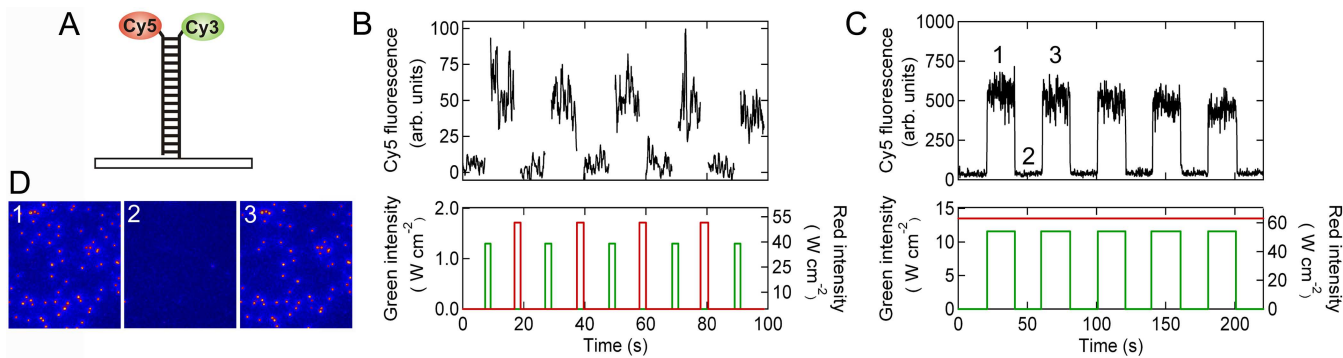}
\caption{\label{fig1}A single-molecule optical switch. (A) Schematic diagram of the single-molecule switch. (B) Lower panel: Alternating red and green laser pulses were used to switch the Cy5 molecule off and on, respectively. A weak, red probe beam (5 W cm$^{-2}$) was constantly on (not shown). Upper panel: the fluorescence time trace of a single Cy5 shows that the molecule was switched between its two fluorescence states. (C) Lower panel: The red laser was continuously on, and the green laser was periodically turned on and off, effectively gating the fluorescence output of the Cy5. Upper panel: the fluorescence time trace of a single Cy5 shows the molecule being switched on and off by the green laser. (D) Fluorescence images of individual Cy5 molecules taken at times specified  in (C). A video showing the switching behaviour is available online [32].}
\end{figure*}

In this work, we demonstrate a single-molecule optical switch which can function as a short-range spectroscopic ruler to probe distances down to 1 nm.  The molecular switch is based on a cyanine dye (Cy5) which we designate as the primary switch molecule, and a secondary chromophore (Cy3) which facilitates switching of the Cy5.  For the purpose of single-molecule detection, the Cy5 and Cy3 were attached to opposite strands of a double-stranded DNA molecule and immobilized on a fused quartz surface through a streptavidin-biotin linkage (Fig.~\ref{fig1}A). Molecules immobilized on a polyethylene glycol (PEG)-coated surface yielded similar results.  Individual Cy5/DNA/Cy3 constructs were imaged in Tris buffer (pH 7.5) with an oxygen scavenging system to reduce the photobleaching rate of the dyes.  Imaging was done using a total internal reflection microscope. For details of materials and methods, see Supplementary materials \footnote{See EPAPS Document No. [] for details of materials and methods used, the kinetics of the switching behavior, and an online video showing the switching of individual molecules between fluorescent and dark states. A direct link to this document may be found in the online article's HTML reference section. The document may also be reached via the EPAPS homepage (http://www.aip.org/pubservs/epaps.html) or from ftp.aip.org in the directory /epaps/.}.  

To demonstrate that this construct functions as a single-molecule optical switch \cite{kulzer97,dickson97,irie02}, we excited it with alternating red (638 nm) and green (532 nm) laser pulses (Fig.~\ref{fig1}B). The fluorescence state of Cy5 was probed continuously by an additional low-intensity red laser beam (638 nm).  Remarkably, individual Cy5 molecules were efficiently switched between two states: the red pulse converted Cy5 to a dark state (designated ``off''), whereas the green pulse reliably brought the molecule back to the fluorescent state (``on'').  Another way to demonstrate the switching effect is by showing that the green laser light effectively ``gates'' the red fluorescence of the Cy5.  Using an excitation scheme involving a single red laser that was continuously on and a green laser that was turned on and off periodically, we were able to switch individual Cy5 molecules between the ``on'' and ``off'' states in a digital manner over multiple cycles (Fig.~\ref{fig1}C). When the green laser was off, the red laser light converted the molecules to the dark state; when the green laser was turned on and operating at intensities sufficiently strong to compete against the action of the red laser, the Cy5 molecules reverted back to the fluorescent state. We note that the observed Cy5 fluorescence was primarily due to direct excitation by the red laser. The contribution arising from FRET between the Cy3 and Cy5 was negligible because of the low green laser intensity used. We note that all of the experiments described below were performed using the ``gating'' excitation scheme.

\begin{figure}
\includegraphics{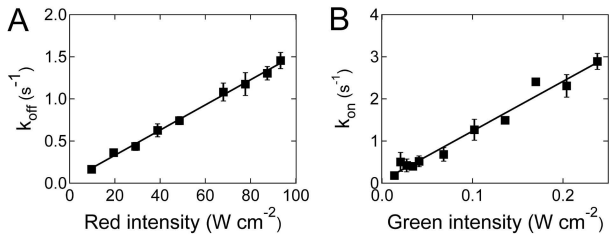}
\caption{\label{fig2} Switching kinetics of single molecules. (A) The rate constant $k_{\mathtt{off}}$ as a function of the red laser intensity. (B) The rate constant $k_{\mathtt{on}}$ as a function of the green laser intensity. Solid lines are linear fits to the data. The standard error for each point is indicated.}
\end{figure}

We characterized the switching kinetics of Cy5 by measuring the number of Cy5 molecules in the fluorescent state as a function of time after the green laser was turned off or on.  We found that the switching behaviour exhibited first order kinetics (see Supplementary materials \footnotemark[\value{footnote}]). Furthermore, the switching rate constants ($k_{\mathtt{off}}$ and $k_{\mathtt{on}}$) each exhibited a linear dependence on laser intensity, indicating that linear optical processes are responsible (Fig.~\ref{fig2}). Switching was rapid compared to other single-molecule optical switches reported \cite{irie02,dickson97}: at laser intensities of 70~W~cm$^{-2}$ (red) and 0.1~W~cm$^{-2}$ (green), molecules can be switched off and on in $\sim 1$ second. We also found the switching to be highly efficient: greater than 90\% of the molecules per field of view could be switched on and off repetitively. The lifetime of the device was limited by photobleaching, typically lasting several minutes under our illumination conditions. We note that the linear intensity dependence may allow the switching kinetics to be extended to faster timescales by using more intense laser excitation.  Also, the lifetime of the device may be dramatically improved by imbedding the molecules in an oxygen-free polymer matrix \cite{english00}.   

\begin{figure}
\includegraphics{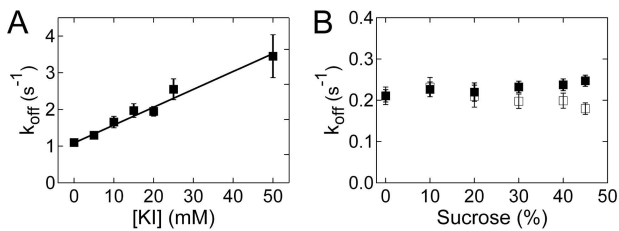}
\caption{\label{fig3} The switching mechanism. (A) $k_{\mathtt{off}}$ as a function of the potassium iodide concentration of the imaging buffer. (B) $k_{\mathtt{off}}$ as a function of the sucrose concentration (w/v). Concentrations of 0\% and 45\% correspond to solvent viscosities of 1 and 6 cP, respectively. The refractive index change of the buffer, induced by sucrose, causes a slight increase in the excitation intensity (amounting to 30\%) at the maximum sucrose concentration used. The $k_{\mathtt{off}}$ values corrected for this effect are shown as open symbols. The changes in the potassium iodide and sucrose concentrations did not affect the absorption coefficient of Cy5.}
\end{figure}

Next we investigated the mechanism responsible for the switching behaviour. Cyanine dyes are known to undergo photo-induced intersystem crossing from an excited singlet state to a dark triplet state \cite{dempster72, chibisov77}. Photo-induced \emph{trans} -- \emph{cis} isomerization of the dye to a non-fluorescent \emph{cis} isomer is also possible \cite{dempster72}. To test these mechanisms, we measured the switching kinetics of Cy5 in the presence of potassium iodide or in solutions of varying viscosity. The rate of intersystem crossing can be strongly enhanced by the presence of iodide due to heavy-atom-induced spin-orbital coupling \cite{kasha52,widengren00}. On the other hand, the \emph{trans} -- \emph{cis} isomerization rate is known to decrease significantly with increasing solvent viscosity due to hindered intramolecular motion \cite{widengren00,sundstrom82,velsko82}. For our system, we found that the rate constant $k_{\mathtt{off}}$ increased linearly with the potassium iodide concentration (Fig.~\ref{fig3}A), but was essentially independent of the solvent viscosity (Fig.~\ref{fig3}B). These results strongly suggest that intersystem crossing mediates the transition from the fluorescent to the dark state. This is consistent with our observation that photo-reversible switching was only observed in buffers with low oxygen content, presumably because oxygen, a known triplet state quencher, increases the spontaneous conversion rate from the triplet state back to the singlet state. 

Investigation of the lifetime of the dark state gave further insight into the switching mechanism.  After 30 minutes without any (red or green) excitation light, only 10\% of the dark Cy5 molecules had returned to the fluorescent state, indicating that the lifetime of the dark state is at least on the order of hours. This lifetime is much longer than previously reported triplet-state lifetimes of cyanine dyes, which were on the order of 100 milliseconds in an oxygen-free environment \cite{english00, widengren00}.  Although we cannot formally rule out the possibility of a long-lived triplet state, our results suggest that the observed dark state is an additional state, subsequent to the triplet intermediate. Potential origins of such a dark state include radical-ion formation and nuclear coordinate movement of the dye \cite{zondervan03,lu97}. Previous studies have reported photoblinking of dye molecules with dark-state lifetimes ranging from milliseconds to tens of seconds \cite{zondervan03}, but none approaching the length of the Cy5 dark state described here.  

The secondary chromophore, Cy3, is essential for the conversion of Cy5 from the dark to the fluorescent state.  In constructs without a Cy3 dye, less than 10\% of dark Cy5 molecules per field of view were switched back on using the green laser at an intensity of 100~W~cm$^{-2}$, three orders of magnitude higher than that required for rapid recovery of Cy5 in the presence of Cy3 (Fig.~\ref{fig2}B). Recovery of Cy5 remained inefficient for other excitation wavelengths that we tested (365, 405, 435, 546 and 577 nm, see Supplementary materials for details \footnotemark[\value{footnote}]).  

\begin{figure}
\includegraphics{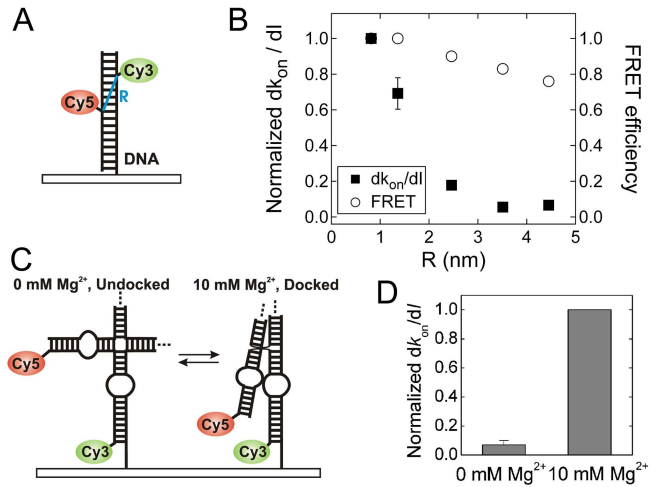}
\caption{\label{fig4} Distance dependence of the Cy3-assisted photo-recovery of Cy5. (A) Schematic diagram of the single-molecule switch with Cy5 and Cy3 separated by a distance $R$. (B) The $R$-dependence of the recovery rate constant of Cy5 $(d k_{\mathtt{on}} / d I)$. The $R$ values are determined according to the structure of B-form double-stranded DNA [32]. We note that the $R$ value reflects the distance between the attachment sites of Cy3 and Cy5 on the DNA but not necessarily the actual distance between the dyes, due to the flexible linkers connecting the dye and the DNA. The $d k_{\mathtt{on}} / d I$ values were obtained from the dependence of $k_{\mathtt{on}}$ on the green laser intensity ($I$), and were normalized against the value obtained at $R$ = 0.8 nm. Also shown are the FRET efficiencies between Cy3 and Cy5, defined as $I_A / (I_A + I_D)$, measured at these distances. $I_A$ and $I_D$ are the fluorescence signals from the Cy5 and Cy3 molecules, respectively, when the Cy3 is excited with 532 nm laser light. (C) Schematic diagram of the hairpin ribozyme in the ``docked'' and ``undocked'' state. (D) Normalized $d k_{\mathtt{on}} / d I$ values for Cy5 on docked and undocked ribozymes.}
\end{figure}

The rate constant for photo-conversion of Cy5 from the dark to the fluorescent state decreased by an order of magnitude as the distance between the Cy3 and Cy5 attachment sites was increased from 1 to 3 nm (Figs.~\ref{fig4}A, B). The observed distance dependence is much steeper than that of FRET between Cy3 and Cy5 or between any dye pairs that have been detected at the single-molecule level. This observation suggests that the switching effect can be used as a spectroscopic ruler to probe distances substantially shorter than what is possible by conventional FRET and may prove useful in investigations of the structural dynamics of biomolecules. 

To demonstrate that the single-molecule switch can report conformational transitions, we applied this technique to the hairpin ribozyme as a model system (Fig.~\ref{fig4}C). The hairpin ribozyme can adopt two conformations: in buffers with high ionic strength, tertiary contacts are formed between the loops A and B (the docked conformation); at low ionic strength, however, the enzyme adopts an undocked conformation where the loop-loop contacts are not formed \cite{walter99,zhuang02,tan03}. We attached Cy3 and Cy5 to the ribozyme at positions such that their separation distance would change significantly when the conformation of the ribozyme changed. The ribozyme was immobilized on a fused quartz surface, and imaged in Tris buffer (see Supplementary materials \footnotemark[\value{footnote}]). As expected, we found that Cy5 molecules on the RNA construct were switched ``on'' and ``off'' in a manner similar to those on the DNA constructs. Significantly, as the magnesium ion concentration in the buffer was raised from 0 to 10 mM, triggering RNA docking, the photo-recovery rate of Cy5 from the dark to the fluorescent state increased dramatically (Fig.~\ref{fig4}D). Using the double stranded DNA samples as a control, we verified that the increase in [Mg$^{2+}$] itself is not responsible for the change in the photo-recovery rate of Cy5. These results indicate that the Cy3-assisted Cy5 switch can readily report the docking/undocking transition of the ribozyme.

The mechanism by which Cy3 facilitates the photo-induced recovery of Cy5 remains unclear. To test whether our observations could be accounted for by long-range charge transfer through double-stranded DNA \cite{kelley97,holmlin98}, we introduced a single AC mismatch in the DNA between the Cy3 and Cy5 attachment sites, a modification which has been shown to reduce the charge-transfer efficiency by 80\% \cite{kelley97}. However, the recovery rate of Cy5 was found to be even greater in such a mismatched construct (by approximately 30\%). Additionally, in our experiments, Cy3 and Cy5 were attached at internal sites of the double-stranded DNA through a linker containing 12 saturated bonds, a geometry that does not promote efficient charge transfer \cite{hurley02}.  Given these considerations, it is unlikely for long-range charge transfer through DNA to be solely responsible for the observed fluorescence recovery. However, we cannot exclude the possibility that other forms of charge transfer are involved. Another process that can potentially explain our observations is energy transfer between the Cy3 and the dark-state Cy5. Short-range energy transfer to a photobleaching intermediate state has been recently reported \cite{ha03}, but it is unclear whether the underlying mechanism is related to the switching behaviour reported here.  

In summary, we have demonstrated a novel single-molecule optical switch and its potential as a short-range spectroscopic ruler. Switching of Cy5 from the fluorescent to the dark state occurs via intersystem crossing, although the remarkably long lifetime of the dark state suggests that it is unlikely to be the triplet state itself but rather an additional state reached via the triplet intermediate. The photo-induced recovery rate depends critically on the secondary chromophore, Cy3, with an intermolecular distance dependence much steeper than that of conventional FRET. Using a model RNA enzyme system, we have shown that this technique can be used to report conformational changes. Continuous measurement of the switching kinetics, under constant green and red laser illumination, should allow this technique to monitor real time conformational changes that occur on the seconds to minutes timescale, a range over which RNA conformational dynamics commonly occur. We expect the time resolution to be further improved by using stronger illumination intensities. A spectroscopic ruler based on this switching effect may significantly improve our capability to probe the structural dynamics of biomolecules by extending the range of length scales measurable with single-molecule techniques.

\begin{acknowledgments}
This work is supported by in part by the ONR, NSF, a Searle Scholarship, a Beckman Young Investigator award, and a Packard Science and Engineering Fellowship (to X.Z.). T.R.B is supported in part by a NIH training grant in Biophysics.
\end{acknowledgments}

\bibliography{bates_et_al_final}

\end{document}